\documentclass{sig-alternate}

\usepackage{subfigure}
\usepackage{graphicx}
\usepackage{epsfig}
\usepackage{color}
\usepackage[ruled,vlined]{algorithm2e}

\begin{document}

%\setcopyright{acmcopyright}
%\CopyrightYear{2017}
%\conferenceinfo{SSDBM '17,}{June 27-29, 2017, Chicago, IL, USA}
%\isbn{978-1-4503-4215-5/16/07}
%\acmPrice{\$15.00}
%\doi{http://dx.doi.org/10.1145/2949689.2949697}
%IF NEED BE.
% --- End of Author Metadata ---

\title{Knowledge Rich Natural Language Queries over Structured Biological Databases}
\numberofauthors{1} %  in this sample file, there are a *total*

\author{
\alignauthor
Hasan M. Jamil\\
       \affaddr{Department of Computer Science}\\
       \affaddr{University of Idaho}\\
       \email{jamil@uidaho.edu}
}

\date{30 July 1999}

\maketitle

\begin{abstract}
Increasingly, keyword, natural language and NoSQL queries are being used for information retrieval from traditional as well as non-traditional databases such as web, document, image, GIS, legal, and health databases. While their popularity are undeniable for obvious reasons, their engineering is far from simple. In most part, semantics and intent preserving mapping of a well understood natural language query expressed over a structured database schema to a structured query language is still a difficult task, and research to tame the complexity is intense. In this paper, we propose a multi-level knowledge-based middleware to facilitate such mappings that separate the conceptual level from the physical level. We augment these multi-level abstractions with a concept reasoner and a query strategy engine to dynamically link arbitrary natural language querying to well defined structured queries. We demonstrate the feasibility of our approach by presenting a Datalog based prototype system, called {\em BioSmart}, that can compute responses to arbitrary natural language queries  over arbitrary databases once a syntactic classification of the natural language query is made.
\end{abstract}



%\vspace*{-2mm}
\section{Introduction}
\label{introduction}

An overwhelming majority of scientific databases use traditional database query interfaces such as SQL and XQuery, or predesigned graphical query interfaces to grant access to their contents. As the information contents of these databases grow more complex in representation, interpretation and analysis, query interfaces needed to access them are also becoming increasingly complicated. Often time, the only convenient method is a predesigned graphical interface through which all accesses are facilitated even though it severely limits the usefulness of these rich information repositories. Although widely used in Life Sciences, these interfaces do not allow ad hoc or spontaneous queries, or investigative queries in ways a natural language interface (NLI) would allow. An encouraging sign is that NLIs are increasingly  being used to provide non traditional query responses in various types of databases such as knowledge repositories \cite{AmsterdamerKM15,Ferre17,DubeyDSHL16s}, text based information repositories \cite{MaioFLP15s,FonteloLA05,Elizabeth08},  biological and clinical databases \cite{SafariP14,HamonGM17,McKnightA12s},  GIS databases \cite{LawrenceR16s} and of course traditional relational databases \cite{SahaFSMMO16,LiJ16,LiPJ14s} with limited scopes.

From the standpoint of end users in Life Sciences, flat view of data is perhaps the most acceptable of all formats while some forms of shallow nesting were also welcomed. Therefore, it can be argued that the relational data model fits well with the practice and end user psychology well in this domain. While XML is creeping its way up in use, querying such data sets using languages such as XQuery or XPath is considered truly difficult still. Therefore, without an abstraction that flattens the nested structure of XML, the natural language processing (NLP) interfaces to such databases \cite{SahaFSMMO16,LiJ16,LiPJ14s} do not appear to have a serious influence. As argued in works such as BioBike \cite{Biobike1,Elhai11}, for over 40 years biologists have voted overwhelmingly not to embrace computer programming as a basic tool through which to look at their world that led to BioBike's development (and many other that followed, e.g., \cite{SafariP14,HamonGM17,McKnightA12s}) that tried to use  NLP as a means to gain access to needed information from the vast array of Life Sciences repositories. But the reality is that such a high level interface still remains illusive mostly because of the representation hurdles, highly interpretive nature of Life Sciences data\footnote{Tool applications are essential before an understanding can be gained for most biological data such as DNA sequences, protein structure, or pathways. Simple read off of the data as in relational model does not reveal any information in general.}, translation from natural language to database queries, and the inherent difficulty in processing NLP queries. A practical approach, we argue, is providing a flat relational view of complex data such as XML so that users can comprehend and query the information repositories at their disposal in the well understood model of flat relations as a middle ground\footnote{This is not to say that other data models are not useful, but to emphasize that flat relations make it possible to use powerful deductive query engines such as Prolog, XSB \cite{XSB} and F-Logic \cite{F-Logic} without much effort as we demonstrate in this paper. In fact, the ORAKEL system \cite{CimianoHHMS08s} has been developed entirely in F-Logic.}. Once chosen, the question remains, how do we facilitate comprehension of a natural language query in a given context, device a strategy to compute its response, and implement the strategy as a traditional query over structured relations stored in local or remote information repositories?

\begin{figure}[ht]
    \centering{\includegraphics[width=.45\textwidth, height = 3.5in]{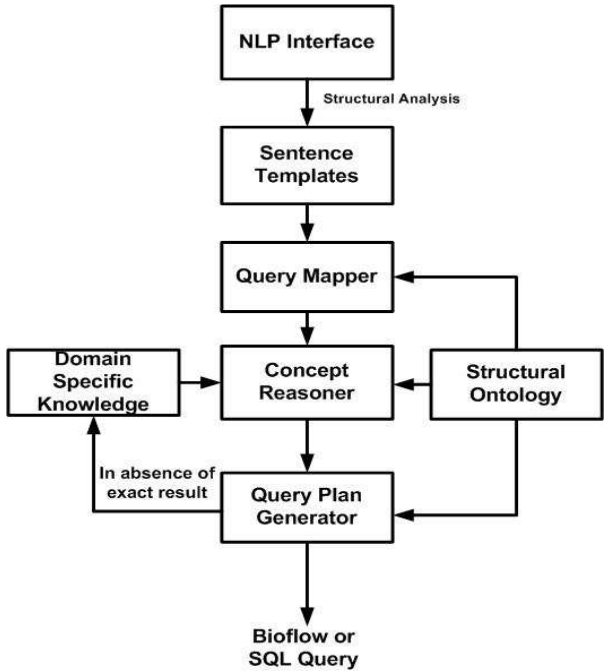}}
    \caption{System components.}
    \label{model}
\end{figure}

In this paper, our goal is to show that a prudent design of multiple abstraction layers to separate the lower level structural details from the upper conceptual level aids in developing a sophisticated reasoning mechanism to analyze the semantic characterization of database objects and their inherent relationships. Once the natural language query's (NLQ) structural characterization is completed, a rule reasoner can be used to analyze the layered representation to map the NLQ to a structured query that not only implements the intent of the upper level query, it does so in a cooperative manner that no traditional query languages such as SQL or XQuery do. By that we mean, we can now actually respond to queries in much the same way a human might do and thereby reflect a deeper semantic understanding of the queries and their true intents. The diagram in figure \ref{model} shows a bird's eye view of the interrelationships between the components in different abstraction layers, the machineries involved, and the conceptual model how the entire system functions to produce near human response to natural language queries.

As a next logical step, we believe it is possible to extend this system toward a cooperative information repository that is capable of sensing the context of a query and carrying the context over an entire investigative session. To that end, we also envision a system where concepts are learned in a query session, and the interrelationship among them and the constraints they impose can be made to influence the candidate set of responses. In an early research on SmartBase \cite{SmartBase-fqas-2009s}, we have shown that non-monotonic inheritance of constraints (SQL's {\tt where} clause conditions) related to concepts in a query to the next query as a logical conjunct simulates the possible world semantics of databases \cite{Hintikka1962,Kripke1963} in a practical and computable way yet offers a better and richer approach to interactive query processing and cooperative response than systems such as \cite{NandiJ09,LiYJ08,mystiq} or earlier research on cooperative query processing \cite{Chu93s,GMN94e,GS96}. We claim that contextual and knowledge rich NLP query processing remains as an extremely difficult proposition in contemporary approaches and the NLP model we propose in this paper can be adapted to include SmartBase query processing features to support a truly contextual NLQ interface.

\begin{figure*}[ht!]
\centering
\subfigure[Protein Table ``UniProt"]{\label{ta}
\begin{tabular}{|l|l|l|}
    \hline
    {\bf ProteinName} & {\bf ProteinID} & {\bf Function}\\
    \hline
    Putative replication & O85067 & Plasmid maintenance \\
    \hline
    Ubiquinol-cytochrome c reductase & Q9NVA1 & Cytoplasmic vesicle \\
    \hline
    F-box domain protein 2 & Q60584 & Substrate recognition\\
    \hline
\end{tabular}}\\
\subfigure[Gene Table ``Entrez"]{\label{tb}
\begin{tabular}{|l|l|l|l|c|}
    \hline
    {\bf GeneName} & {\bf GeneID} & {\bf UniProtProteinID} & {\bf DNASequence} & ... \\
    \hline
    FBXW2 & 30050 & Q60584 & CTCTTTCTTTCG ... & ...\\
    \hline
    repA1 & 1246500 & O85067& ACCCTTGGAAACCC... & ...\\
    \hline
    FBXW2 & 26190 & Q9UKT8 & CTCTTTCTTTCT ... & ...\\
    \hline
    UQCC & 55245 & Q9NVA1& TTTTGGGGCCCCAAA... & ...\\
    \hline
\end{tabular}}\\
\subfigure[Query 1 Response]{\label{tc}
\begin{tabular}{|l|}
    \hline
    {\bf DNASequence}\\
    \hline
    CTCTTTCTTTCG ...\\
    \hline
\end{tabular}}
\subfigure[Query 2 Response]{\label{td}
\begin{tabular}{|l|}
    \hline
    {\bf Function}\\
    \hline
    Cytoplasmic vesicle\\
    \hline
\end{tabular}}
\subfigure[Enhanced Response 1]{\label{te}
\begin{tabular}{|l|}
    \hline
    {\bf DNASequence}\\
    \hline
    CTCTTTCTTTCG ...\\
    \hline
    CTCTTTCTTTCT ...\\
    \hline
\end{tabular}}
\subfigure[Enhanced Response 2]{\label{tf}
\begin{tabular}{|l|}
    \hline
    {\bf Function}\\
    \hline
    Cytoplasmic vesicle\\
    \hline
    Substrate recognition ~~~\\
    \hline
\end{tabular}}
\caption{Queries over UniProt and Entrez databases.}
\label{tables}
\end{figure*}

\subsection{A Motivating Example}
\label{motexample}

Consider a protein/gene function database consisting of two base tables -- {\em UniProt Protein Table} and {\em Entrez Gene Table} as shown in figures \ref{ta} and \ref{tb} respectively. Now consider answering queries of the form
\begin{enumerate}
\item {\em ``List all F-box domain protein 2 sequences"}, or
\item {\em ``What are the functions of UniProt proteins Q9UKT8 and Q9NVA1"}?
\end{enumerate}
In a conventional relational database, these two natural language queries will most likely map to the following to two SQL queries to return the answer sets shown in figures \ref{tc} and \ref{td} respectively.

\begin{verbatim}
select DNASequence
from Entrez as e, UniProt as u
where u.ProteinID=e.UniProtProteinID and
  ProteinName="F-box domain protein 2"

select Function
from UniProt
where ProteinID in {Q9UKT8, Q9NVA1}
\end{verbatim}

The scientific fact here is that the UniProt protein {\tt Q60584}, with corresponding gene name {\tt FBXW2}, is a mouse protein. Functionally,  {\tt Q60584} is identical to UniProt protein {\tt Q9UKT8} and have the same gene name, {\tt FBXW2}, is a human protein. Furthermore, they also have almost identical DNA sequences. This similarity can be computed -- (i) by homology of the two sequences corresponding to the UniProt IDs {\tt Q60584} and {\tt Q9UKT8} as shown in Entrez Gene Table, (ii) knowing that these two genes are orthologs, (iii) by determining that they have identical gene names in Entrez Gene Table, (iv) by consulting a ID mapping database such as GeneCard \cite{GeneCard} and finding their equivalence, or (v) by consulting the Gene Ontology (GO) database \cite{GO}, among many other ways. If such background knowledge is used, then it becomes possible to return the tables shown in figures \ref{te} and \ref{tf} with an additional response in each table, as knowledge derived responses. Notice that such inferences are not directly possible in traditional relational databases, but these responses are valid, and useful. The BioBike system actually takes this approach.

In contrast, in BioBike, users must specifically perform a homolog computation in order to receive the responses shown in figures \ref{te} and \ref{tf}. The system only provides the computational means. The novelty of our system is that such knowledge is derived by the system on its own and no explicit requests need to be made. In a more recent work \cite{ConLog-bibm-2011s,ConLog-bmcgenomics-2012}, it was shown that an ontology based logical query language ConLog can be used to non-intrusively capture semantic knowledge and query response generated without such explicit knowledge entailed response computation. In this paper, our goal is to show that a canonical representation, as shown in figures \ref{ra} through \ref{re}, can be used to compose queries in a logic based language such as Datalog without user intervention. More importantly, we show that natural language sentences can be classified into sentence structure templates, and these templates can be used to select appropriate query predicates to fire to compute intended responses. In the next few sections, we present a detailed discission how our approach works.

\begin{figure*}[ht!]
\centering
\subfigure[Derivatives: der]{\label{ra}
\begin{tabular}{|l|l|}
    \hline
        {\bf ConceptA} & {\bf ConceptB} \\
        \hline
        Gene & Protein \\
    \hline
\end{tabular}}
\subfigure[Foreign Keys: forK]{\label{rf}
\begin{tabular}{|l|l|l|l|}
    \hline
    {\bf TableX} & {\bf TableY} & {\bf ColumnX} & {\bf ColumnY}\\
    \hline
    Entrez & UniProt & UniProtProteinID & ProteinID \\
    \hline
\end{tabular}}
\subfigure[Similar Concepts: simCon]{\label{rc}
\begin{tabular}{|l|l|l|}
    \hline
    {\bf ConceptX} & {\bf ConceptY} & {\bf Relation}\\
    \hline
    Gene & Gene & Ortholog, Paralog, Duplication \\
    \hline
\end{tabular}}
\subfigure[Tools: cTool]{\label{rb}
\begin{tabular}{|l|l|}
    \hline
    {\bf Relation} & {\bf Operation}\\
    \hline
    Ortholog & BLAST, ORSCAN\\
    \hline
    Paralog & GENCODE \\
    \hline
\end{tabular}}
\subfigure[Canonical Database: canDB]{\label{rd}
\begin{tabular}{|l|l|l|l|}
    \hline
        {\bf Concept} & {\bf PrimaryKey} & {\bf AttributeName} & {\bf AttributeValue}\\
        \hline
        Gene & 1246500 & GeneName & repA1\\
        \hline
        Gene & 1246500 & UniProtProteinID & O85067\\
        \hline
        Gene & 1246500 & DNASequence & ACCCTTGGAAACCC...\\
        \hline
        Gene & 1246500 & GeneID & 1246500\\
        \hline
        Gene & 55245 & GeneName & UQCC\\
        \hline
        Gene & 55245 & UniProtProteinID & Q9NVA1\\
        \hline
        Gene & 55245 & DNASequence & TTTTGGGGCCCCAAA...\\
        \hline
        Gene & 55245 & GeneID & 55245\\
        \hline
        Protein & O85067 & ProteinID & O85067\\
        \hline
        Protein & O85067 & Function & Plasmid maintenance\\
        \hline
        Protein & O85067 & ProteinName & Putative replication\\
        \hline
        Protein & Q9NVA1 & Protein ID & Q9NVA1\\
        \hline
        Protein & Q9NVA1 & Function & Cytoplasmic vesicle\\
        \hline
        Protein & Q9NVA1 & ProteinName & Ubiquinol-cytochrome c reductase\\
        \hline
        ... & ... & ... & ...\\
        \hline
\end{tabular}}
\subfigure[Structural Ontology (sO)]{\label{re}
\begin{tabular}{|l|l|l|l|}
    \hline
    {\bf ConceptName} & {\bf TableName} & {\bf Key} & {\bf Attributes}\\
    \hline
    Gene & Entrez & GeneID & GeneName, DNASequence, ...\\
    \hline
    Protein & UniProt & ProteinID & Function, ProteinName, ...\\
    \hline
\end{tabular}}
\caption{Internal representations and meta data for knowledge rich queries.}
\label{tables2}
\end{figure*}

\subsection{Related Research}

Our research in this paper is mainly influenced by the BioBike project \cite{BioBike2,Biobike1} at Virginia Commonwealth University. In our own lab, we are focused on developing smart declarative querying and workflow engines for distributed and heterogenous scientific databases. We have developed a declarative query language for data integration called BioFlow \cite{BioFlow-swf-2009s,BioFlow-ijbpim-2010s} based on which we have developed two visual interfaces called VizBuilder \cite{VizBuilder-Jamil-TCBB-2013s, VizBuilder-dils-2009s} and VisFlow \cite{VisFlow-IJDMB-2017s,MouJM17s,VisFlow-BIBM-2016s} to support distributed workflow queries for Life Sciences data management and querying system called the LifeDB \cite{LifeDB-dexa-2009s}. To complement LifeDB, we have also develop an ID mapping database called the MapBase \cite{MapBase-TCBB-2015s} and phylogenetic database PhyloBase \cite{PhyloBase-TCBB-2017s,PhyloBase-SSDBM-2016s}.

The promise of knowledge rich query interfaces in Life Sciences inspired us to explore if systems could detect query intent, especially in exploratory settings such as in SmartBase in which we developed a contextual query processor capable of recognizing the query skyline \cite{KossmannRR02s} from a set of successive queries. Augmenting structured databases with semantic knowledge for the purpose of knowledge rich queries have been studied in the context of NLQs \cite{ConLog-bibm-2011s,ConLog-bmcgenomics-2012}, and structured database integration \cite{Jamil14s}. Despite many such efforts, articulating application description by domain scientists in Biology to facilitate knowledge rich queries remain a significant challenge. This, we believe, is mainly due to a complex regiment of interpretive tools needed to analyze the raw data, the domain specific knowledge that leads to many alternative and approximate response to a query, and so on. From these standpoints an NLP interface seems to be very attractive for this community and BioBike stands out as an interesting model as it demonstrated its ability to respond to Biological queries in an intuitive way. In this paper, our goal is to fuse these two approaches to be able to respond to natural language queries over arbitrary biological databases.

While interesting in its approach, BioBike still seems to have encoded the query processing logic into the system in a way that did not support good enough abstractions that made it a less likely candidate for a viable architecture that designers could just adopt as a generalized engine for widespread use. To a large extent, it still remained a seriously hard wired programming application. Though NLP interfaces for traditional structured databases are extensively studied in the literature (e.g., \cite{SahaFSMMO16,LiJ16,LiPJ14s}), knowledge rich queries using NLP \cite{JosephA91s,KaufmannB10} have been rarely and tangentially explored. Mostly existing NLP systems map NLQs to structured queries only when the database structures, application domain and the interpretation of the data are known and well understood. Although NLP is being increasingly used to return possible answers \cite{CookPCMHB16,AletrasTPL16,GkatziaLR16}, we stand out in the way we conceptualize the database, represent them in a canonical form, and allow an intelligent and user specific analysis of the content of the database to map the NLQ to a target declarative query language. We are unaware of any system that follow this approach including BioBike.

The remainder of the paper is organized as follows. In section \ref{motexample}, we have already presented an overview of the system and discussed its various components on intuitive grounds. Then to illustrate the functionalities of our system, we discuss several expository examples and their execution in reference to the system components just presented in section \ref{system}. In contrast with many NLP based query processing system, we aim to support description based query processing in which a set of sentences together describe a complete workflow of an application in the direction of SmartBase. The type of NLP queries we support are introduced in section \ref{queryt} and demonstrate that the type of queries we support covers a wide class of queries, and they can be nested to pose more semantically rich queries. Our goal is to build a middleware that is capable of interpreting the problem description by recognizing the intent of the overall workflow, and offer a near NLP experience and return a more cooperative response than traditional databases. From this standpoint, we also highlight the salient features of our system that distinguishes it from the leading contemporary systems. In section \ref{knowRQ}, we discuss the heart of our system BioSmart, its concept reasoner that helps compute knowledge rich queries using domain specific knowledge. The method to incorporate application specific desktop or online computational tools in Datalog are also discussed since custom designed and generic computational tools play a significant role in biological data processing. In section \ref{XSB}, we discuss Java specific calls to facilitate invocation of computational tools from XSB, the reasoning platform we use in BioSmart. In section \ref{access}, we discuss a more generic and more powerful approach to modeling external function calls in the form of a database engine using workflows. Finally we summarize and discuss possible future research in section \ref{conclusion}.

\section{BioSmart System Overview}
\label{system}

Representation and interpretation of biological data are complex, and analyzing them requires domain specific knowledge which may vary from user to user. To support such a fluid and knowledge driven querying environment, we offer a free-from NLP interface called {\em BioSmart}. In this interface, users may ask arbitrary queries that are parsable as a valid natural language sentence. We then categorize these sentences into several classes that fit into predefined syntactic templates\footnote{We defer the discussion on the specific mapping algorithm from natural language to categorized templates in this paper for the sake of brevity. Instead, we mainly focus on how such templates once generated are interpreted to respond to queries.}. These templates are then analyzed in the context of the underlying database structures and and assigned an interpretation entailed by its logical meaning. We expect the analysis to yield a query and a query execution plan in a declarative language such as SQL, Datalog, or BioFlow.

\subsection{Components and Overview of the System}
\label{Med4}

Let us introduce the architecture of the NLQ engine we envision with the help of an example. Consider the query $Q$ below.
\begin{quote}
{\em ``Find the photosynthetic genes of cyanobacteria Prochlorococcus sp. strain (known also as MED4)".}
\end{quote}
Prochlorococcus is an extremely small Chl {\em b}-containing  light-harvesting cyanobacterium antenna system sometimes constituting up to 50\% of the photosynthetic biomass in the oceans \cite{HessRTLSLC2001}. So, the query above is interesting at many levels. First, although there is a fairly recent database on cyanobacteria called {\em CKB} \cite{PeterLMVTKDGL2015}, it is not so simple to dig out the information this query seeks from this database. The search term Prochlorococcus does not pull up any information from CKB. But, from the literature \cite{HessRTLSLC2001} we know that Prochlorococcus has two strains, {\tt MED4} and {\tt MIT9313} that are representatives of high and
low-light adapted ecotypes, characterized by their low or high Chl {\em b/a} ratio, respectively. Furthermore, {\tt MED4} is more recently evolved and has about 1,686 protein coding genes while {\tt MIT9313} belongs to the
most deeply branching lineage of Prochlorococcus with 2,200 genes. The Prochlorophyte Chlorophyll-Binding proteins (PCBs) responsible for photosynthesis in Prochlorococcus are encoded by
a single gene in all the low {\em b/a} strains, whereas multigene families have been
found in several high b/a strains. It is also known that {\tt pcb} is a gene in the high light-adapted {\tt MED4}, and {\tt pcb1} and {\tt pcb2} are two genes in low light-adapted {\tt MIT9313} strain for photosynthesis.

However, it is not possible to isolate these genes from CKB using {\tt pcb} as the search term to know if their function is photosynthesis. To discover this knowledge, one must sift through to select all rows having {\tt pcb} as a column, and link the gene names to either UniProt of KEGG databases to see if their functions include photosynthesis. Two such gene names are {\tt PMM0627}, and {\tt pcbA} or {\tt Pro0783} that are listed to have photosynthesis as their functions in KEGG and UniProt databases respectively. So the innocent looking query is not that simple at all to compute without knowing all these details from the literature in the first place and thus obviating the need for querying. Even when one wants to learn more, starting off with a wrong database or inappropriate search term may throw the search in a wrong direction. Finally, phylogenetically the closest relative of Prochlorococcus is the organism Synechococcus, and there is great deal of information on this organism that can be leveraged to learn details about photosynthetic genes of Prochlorococcus indirectly. Therefore, we believe the novelty of BioSmart is that given a query $Q$, a database $D$ and a knowledgebase $K$, it computes the response to the query as the entailment relation $D\cup K \models_Q A$, which to our knowledge no other leading databases or query answering system in Life Sciences do.

\subsection{The Generative Process}

The entailment relation we have introduced above is captured in the schematic architecture of BioSmart shown in figure \ref{model}. The natural language sentences in BioSmart are accepted and parsed by its {\em NLP Interface} (NLI) and mapped to predefined sentence or query templates. It is possible that a sentence will be broken down into several such templates that capture the meaning. The {\em Query Mapper} (QM) component transforms the templates into a logical query using the information in the {\em Structural Ontology} (SO) in the context of the underlying database -- identifies the tables, the analysis tools needed, and the possible joins needed to compute the query.

In the SO, we maintain concepts and their properties independent of their structural affiliation in tables or their specificity as objects in a table. For example, the two tables {\em Entrez} Gene Table and {\em UniProt} Protein Table in figures \ref{ta} and \ref{tb} will be described in SO as a meta-data table as shown in figure \ref{re}. The concepts in SO (e.g., Gene and Protein in column {\em ConceptName}) can now be linked to natural language concepts independent of their table affiliation, which now can actually be discovered from the table in figure \ref{re} using column name {\em TableName}. Other components such as {\em Query Mapper {\em (QM)}, Concept Reasoner} (CR) and {\em Query Plan Generator} (QPG) also use the services by and the information maintained in SO.

\subsection{Query Mapping}

The CR subsystem uses a set of axioms to discover how responses can be generated using the objects in the database for a given natural language sentence templates. This is the component that also discovers relationships that can be used to construct alternative responses, either directly or indirectly. It also discovers any need for tool applications to find query responses. The interesting aspect of this component is that the axioms used in CR are not domain specific, and so, the system can be used for other domains as is. However, {\em domain/concept specific knowledge} (DSK) in the form of a knowledge base can be supplied as a plug-in to tailor the functionality and bring specificity to the system. Better or more richer the knowledge-base is, more sophisticated response it is capable of generating. The union of DSK, CR and SO serves as the knowledgebase $K$ in the entailment relation $D\cup K \models_Q A$, and the response $A$ is as rich as entailed by $K$.

The query specification generated by the CR is then forwarded to the QPG which in consultation with the SO transforms the specification into a set of database specific executable queries in a language of choice. QPG achieves this goal by using the \emph{canonical representation} of the existing data sources. In the canonical representation, we represent the underlying database in a triple form similar to RDF \cite{WSDL:URL}. All the tables from the user-specified database are broken up in a $\langle${\em concept, attribute type, attribute value}$\rangle$ triple format. Concept is a combination of the concept types and the primary key for a conceptual object. Attribute type is the column name from the original data table and attribute value is the value stored in that column for a particular concept. Canonical representation of the data sources shown in the figures \ref{ta} and \ref{tb} is presented in figure \ref{rd}. The steps involved in the mapping process is outlined in algorithm \ref{alg:map} at a very high level.

\IncMargin{1em}
\begin{algorithm}[h]
%\SetLine
{\footnotesize
\KwIn{A natural language query $Q$}
\KwOut{Executable declarative query $Q_e$}
Perform syntactic analysis of $Q$ to generate a parse tree $T$\;
Perform structural analysis of $T$ to match a sentence template $s\in S$\;
Generate logical equivalent $L$ of $s$\;
Apply DSK to $L$ to generate conceptual plan $C$\;
Generate executable script $Q_e$ from $C$\;
Execute $Q_e$\;
\If{$Q_e$ succeeds}{
Return result $A$\;
Exit\;
}
\Else{
Try alternate mapping, if possible\;
}
\caption{Mapping NLQ $Q$ to executable knowledge rich query $Q_e$.}
\label{alg:map}
}
\end{algorithm}
\DecMargin{1em}

\section{Query Types}
\label{queryt}

Since interpretation of arbitrary natural language sentences are difficult, keeping in mind that most flat and structured databases use a set based model, natural language sentences that mimic {\tt select-project-join} (SPJ) queries are our priority. Such queries have limited interpretive scopes in terms of what they allow. For example, queries are mostly about objects and their properties, and their relationships with other objects. Often, we need to construct complex objects by piecing together parts from various tables. We believe almost all such query structures can be admitted if we recognized three basic types of sentence structures -- {\em iterative}, {\em conditional}, and {\em imperative} or {\em interrogative} natural language queries. Keeping the admissible sentence structures set simple and small, we actually follow SQL's footsteps in which complex and more expressive queries can be built by nesting simple structures arbitrarily. Thus, most of the biological queries potentially can be expressed by one of these types, or by a combination of them. In BioSmart, we do so by allowing series of queries in succession in a context.

\subsection{Iterative Queries}

\begin{figure}[ht!]
\centering
    \subfigure[Iterative]{\label{pt1} \includegraphics[height=1in,
    width=.23\textwidth]{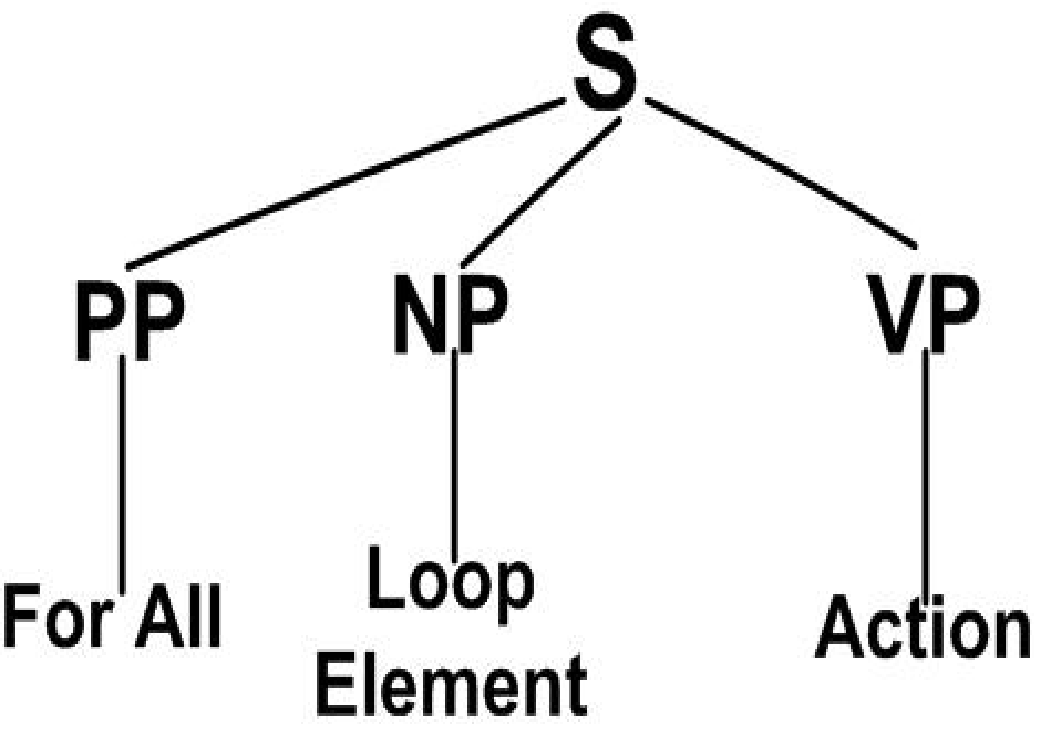}}
    \subfigure[Conditional]{\label{pt2} \includegraphics[height=1.25in, width=.23\textwidth]{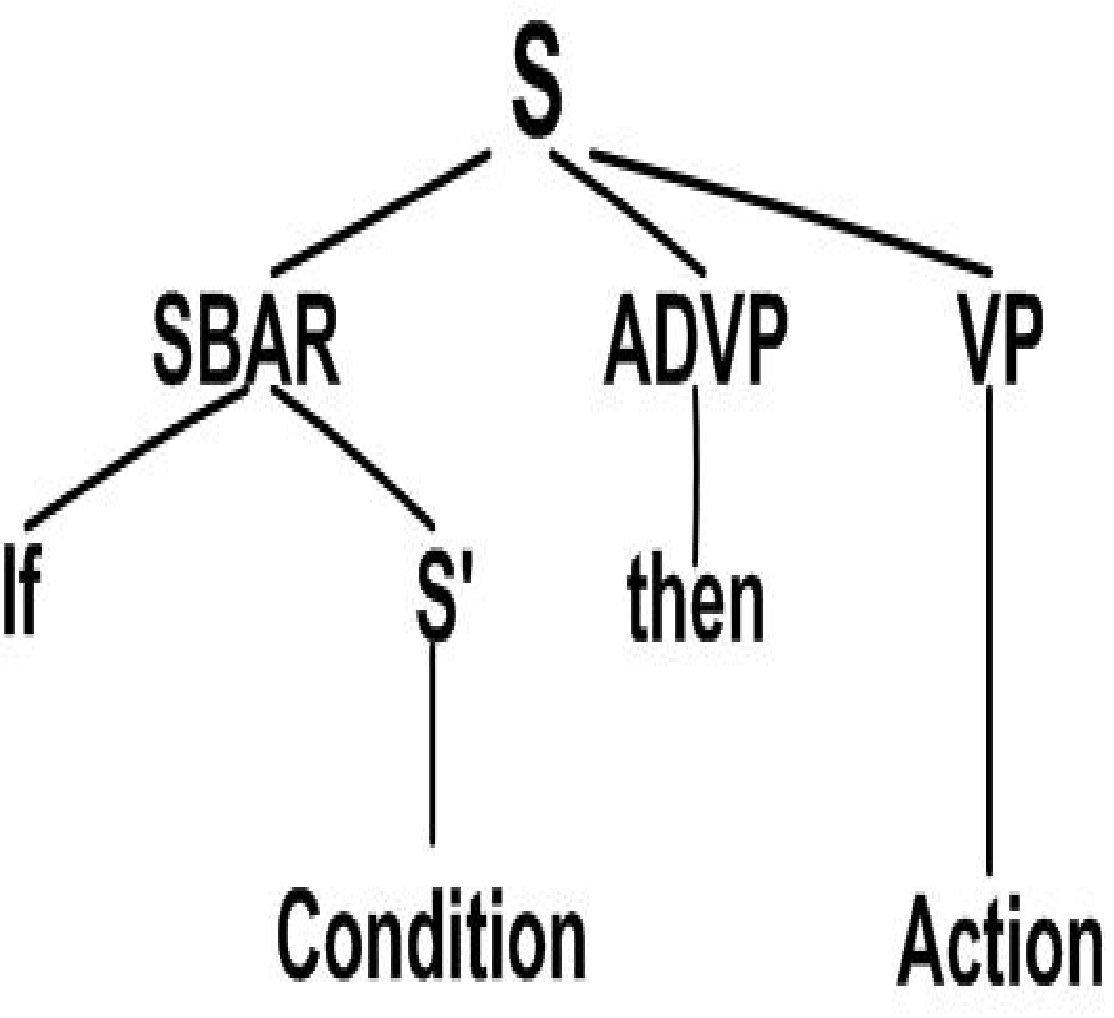}}
    \subfigure[Imperative]{\label{pt3} \includegraphics[height=1in, width=.175\textwidth]{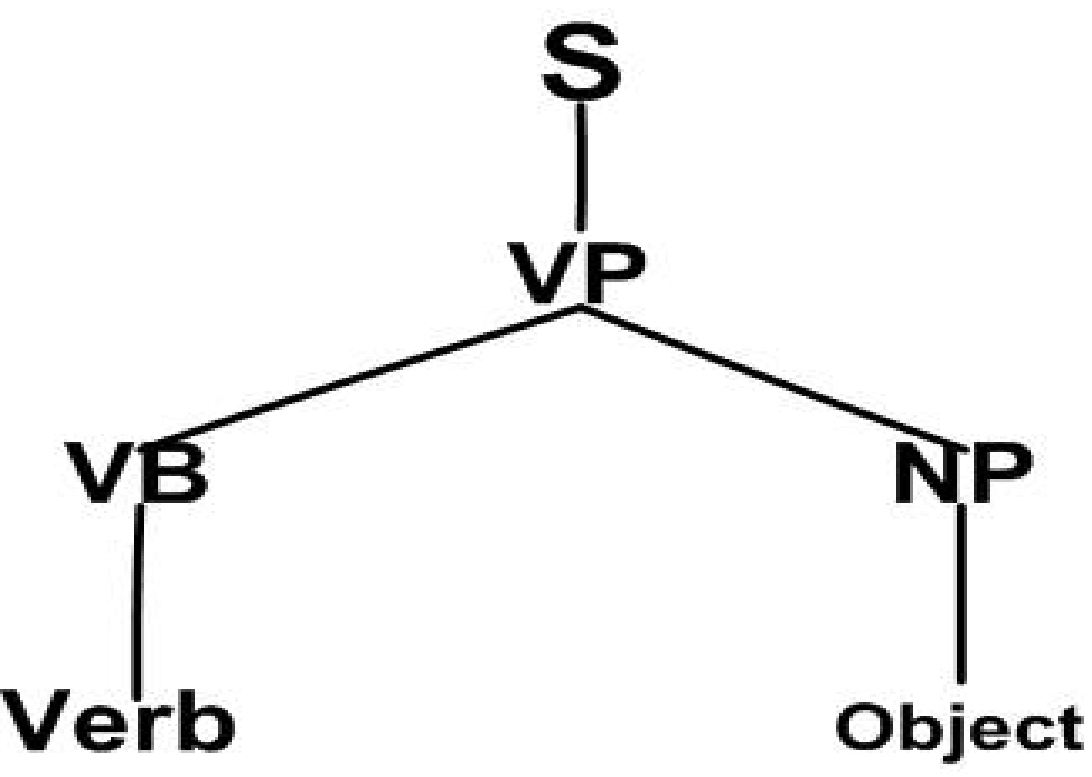}}
    \subfigure[The noun phrase structure]{\label{pt4} \includegraphics[height=1.25in, width=.25\textwidth]{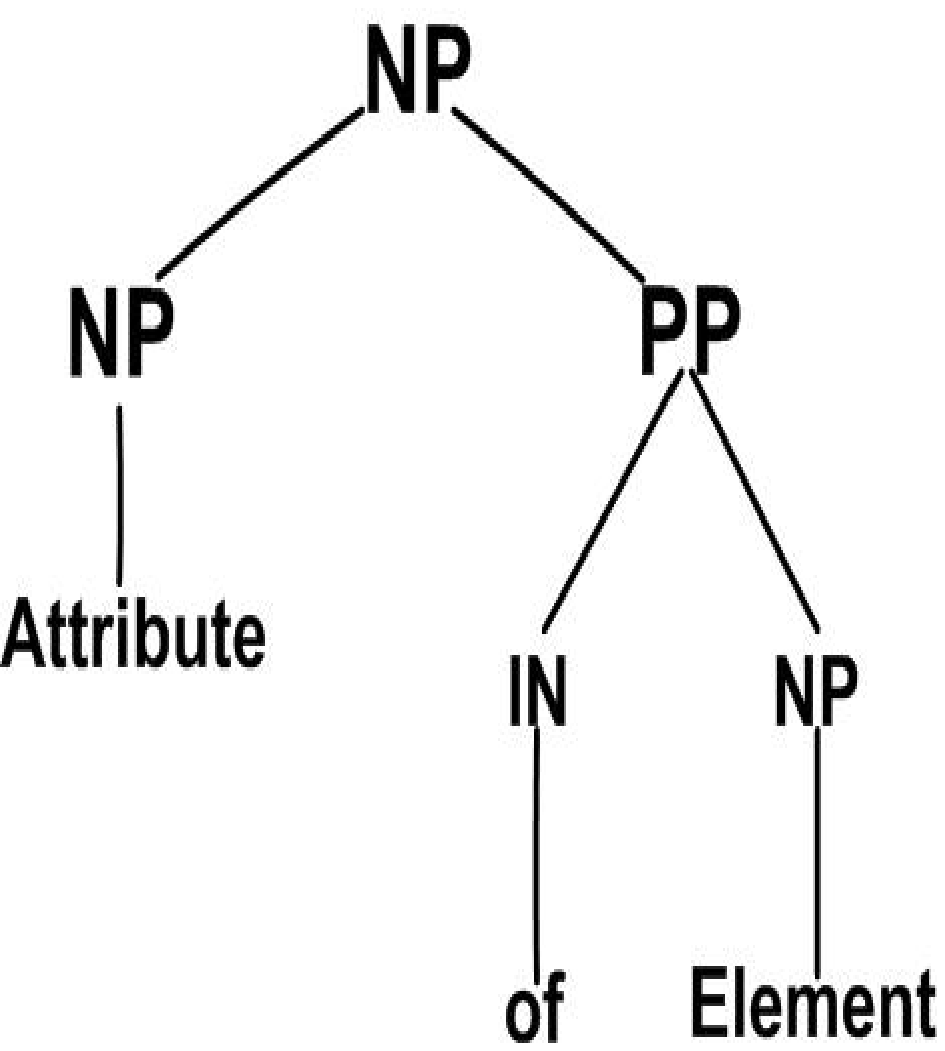}}
    \caption{Natural language query templates.}
\label{templates}
\end{figure}

Consider expressing a query in English in two different but semantically equivalent ways.
\begin{quote}
{\em ``List the functions of all human genes"}
\end{quote}
and,
\begin{quote}
{\em ``For all Homo sapiens genes, list their functions."}
\end{quote}
These two seemingly two structurally different queries in English actually map identical structured queries in SQL, or at least can be expressed by a single query from table \ref{ta}. But as a query type, these natural language queries ask to retrieve objects that satisfy certain properties (including empty properties). If we parse these queries in a syntax tree as shown in figure \ref{templates}, we will discover that each of these query types roughly adhere to one of the parse tree structures in this figure, we call sentence templates. The queries above can be parsed to resemble the template shown in figure \ref{pt1}, or the iterative type.

\begin{figure}[ht!]
\centering
    \subfigure[Iterative]{\label{pt5} \includegraphics[height=1.75in, width=.4\textwidth]{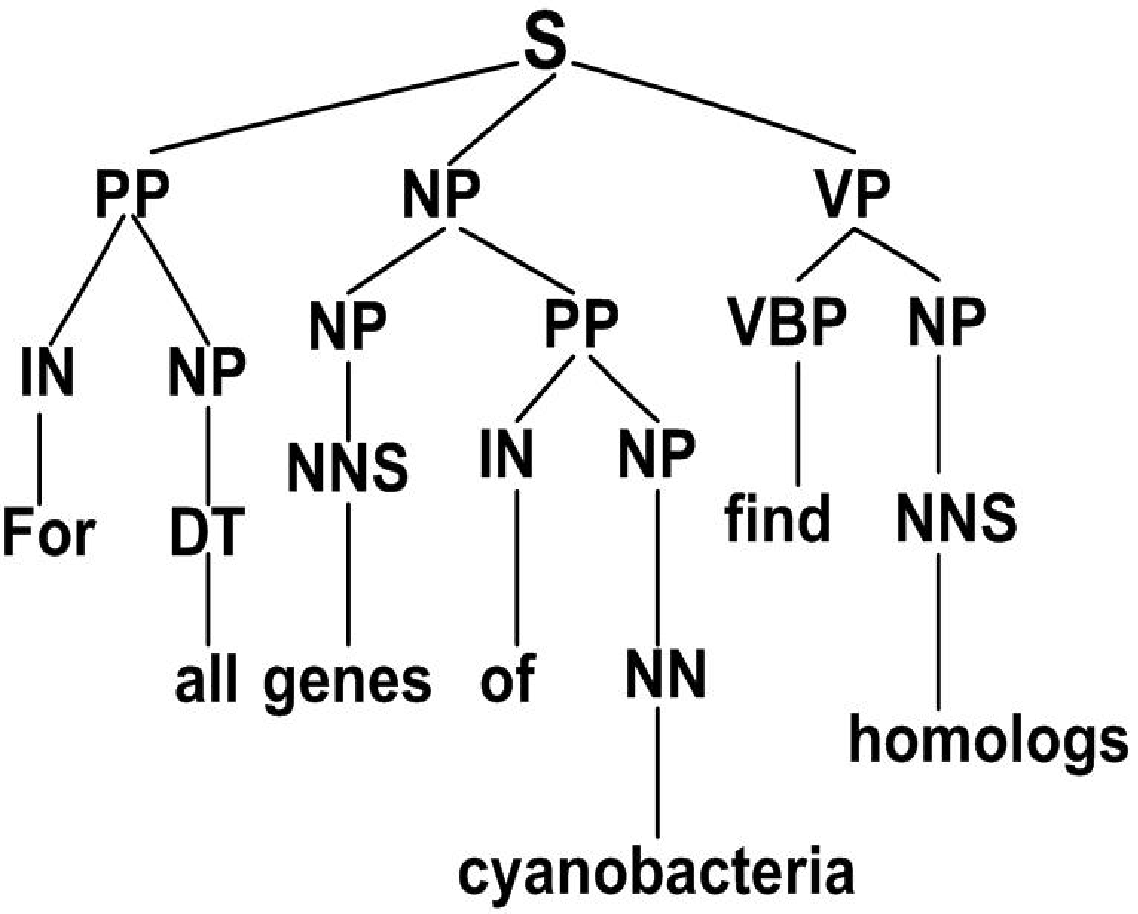}}
    \subfigure[Conditional]{\label{pt6} \includegraphics[height=2.25in, width=.4\textwidth]{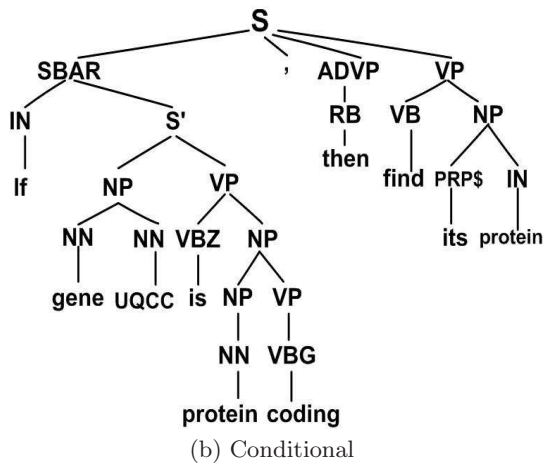}}
    \subfigure[Imperative]{\label{pt7} \includegraphics[height=2in, width=.3\textwidth]{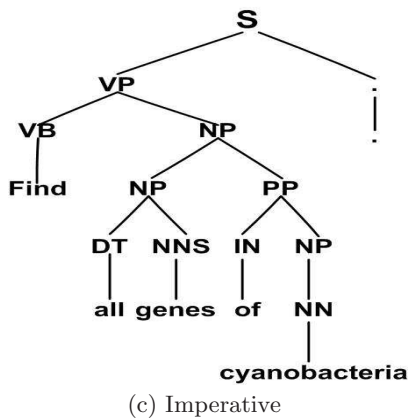}}
    \caption{Natural language query examples.}
\label{templatesExam}
\end{figure}

An iterative, or {\em loop}, type natural language query, and thus its template, starts with a prepositional phrase (PP), followed by a noun phrase (NP) and then a verb phrase (VP). The prepositional phrase usually includes the phrase ``for all", or its variants. The noun phrase essentially points to the objects or entities we are to apply the verb phrase, i.e., properties or actions. The parse tree for the query
\begin{quote}
{\em ``For all genes of cyanobacteria find homologs"}
 \end{quote}
is shown in figure \ref{pt5} which actually is an instance of the iterative query template in figure \ref{pt1}. In this query the NP is {\em all genes of cyanobacteria} and the VP is {\em find homologs}.

\subsection{Conditional Queries}

Iterative queries are not required to satisfy any constraint. In other words, they can be viewed as a simple projection query in SQL (i.e., {\tt select from}), possibly with a simple {\tt where} clause condition for the purpose of object or property identification. In contrast, a {\em conditional} query specifies an arbitrary precondition that the objects must satisfy in the form of an {\em if then} structure as shown in the query template in figure \ref{pt2}. In such queries, a NP-VP sequence follows the {\em if} condition (i.e., the $S'$ subtree), and the VP following the NP captures the action clause, where the NP-VP sequence has a simple {\em subject-verb-object} sentence structure. The VP after the {\em then} has a {\em verb-object} imperative sentence structure. Figure  \ref{pt6} shows the parse tree for the query
\begin{quote}
{\em ``If gene UQCC is protein coding, then find its protein"},
\end{quote}
as an instance of the template in figure \ref{pt2}. In this example {\em ``is protein coding"} is the VP for the {\em if} condition, and {\em ``find its protein"} is the VP for the {\em then} inference.

\subsection{Imperative Queries}

{\em Imperative} sentences or queries are basically a verb phrase (VP) consisting of a verb (VB) and an object (NP) on which the verb is to be applied. The template representing an imperative query is shown in figure \ref{pt3}. These type of sentences are used as standalone queries or as a part of more complex iterative of conditional queries. Usually the object (NP) has a structure of the form {\em ``attribute of element"} as in figure \ref{pt4}. As an example, consider the query
\begin{quote}
{\em ``List all genes of cyanobacteria"},
\end{quote}
and its parse tree shown in figure \ref{pt7}. Here the verb is {\em List} and the object is {\em genes of cyanobacteria} in conformance with the structure of figure \ref{pt4}. As discussed in section \ref{Med4}, we will be correct to return genes of Prochlorococcus and Synechococcus, and {\tt MED4} and {\tt MIT9313}, such as {\tt pcb} and {\tt pcbA}.

\section{Knowledge Rich Querying}
\label{knowRQ}

The query identification and processing apparatus we have introduced earlier can now be leveraged to respond to the simple knowledge rich query
\begin{quote}
{\em ``Find the function of gene repA1"}
\end{quote}
in Escherichia coli. Though this query appears simple, as discussed in section \ref{Med4}, finding answer to it may require the use of multiple biological data sources and tools. A user can go to NCBI GenBank to find its function. Failing to find the answer in GenBank, she can find its UniProt ID {\tt P03066} and try to find its function from UniProt database, and a full annotation can also be obtained from GO database. In UniProt, the function for {\tt repA} (Replication initiation protein) is listed as {\tt plasmid replication}, and {\tt copy control}, whereas GO annotation includes \{{\tt DNA replication, plasmid maintenance}\}. In the event none of these databases produced any useful information, she could try to use NCBI BLAST search to find the orthologs of {\tt repA1} and then find the function of those orthologs.

\begin{figure}[h]
    \centering{\includegraphics[width=.35\textwidth, height = 2.25in]{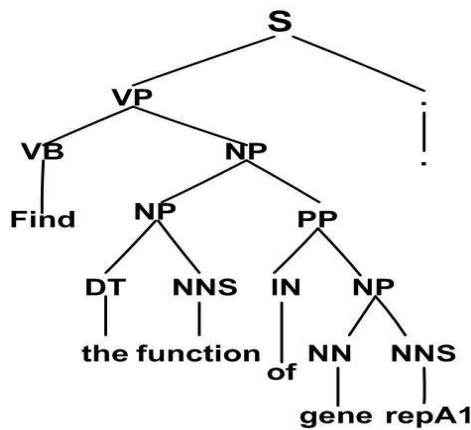}}
    \caption{Parsed tree of example query.}
    \label{exampleQuery}
\end{figure}

Such approaches require a user to be familiar with all these details, and complex queries require even more complex and intricate interwoven knowledge will be essential. Moreover, different resources and tools have unique representation and naming policy which makes it even more difficult to find answers. For example, the GeneCards database assigns {\tt RPA1} as the ID for {\tt repA1}, which is supposed to be an authoritative site for finding the IDs for {\tt repA1} in various databases. Apparently, {\tt repA1} gene symbol has been discontinued and replaced with the name {\tt RPA1}. To aid biologists in such a confusing landscape, in BioSmart, we approach the computation of the query in several steps. We use the Stanford parser to first parse the query submitted to BioSmart NLI and attempt a mapping to one of the query templates in figure \ref{templates} as resulting in the imperative query parse tree in figure  \ref{exampleQuery}, with  action {\em Find} on attribute {\em function} of element {\em gene repA1}.

\subsection{Direct Concept Reconstruction}

From the discussion in section \ref{motexample} it may not have been apparent that we actually do not use the base tables in figure \ref{tables}. Instead, we always use the derived tables in figure \ref{tables2}. In particular, all tables are collapsed into the canonical database form in figure \ref{rd} that allows reconstruction of the objects in any table using an object identifier, the primary key of each table represented as a concept. The remaining set of tables in figure \ref{tables2} forms the DSK and SO components. The reconstruction process depends on the query type, and the complexity of the query and its interpretation assigned.

In the query above, the goal is to compute the {\em function} (a property) of a {\em gene} (a concept/object) named {\em repA1} (attribute value) corresponding to another property (i.e., {\em gene name}). QM uses this interpretation to map the properties ( to the attribute {\em AttributeName}, the concept to the attribute {\em Concept}, and attribute value to {\em AttributeValue}. Essentially, we are trying to compute the pair $\langle${\em function, AttributeValue}$\rangle$ for the concept {\em gene} having $\langle${\em GeneName, repA1}$\rangle$ as an entry in the canonical database {\em CanDB}. Logically, the response can be constructed by an equi-join of the {\em CanDB} table on the column {\em PrimaryKey} for the {\em Concept} {\em gene}. In general, the reconstruction is captured in the following CR rule,
\begin{verbatim}
1: res(Con, Pk, AttName, AttVal) :-
      canDB(Con, Pk, AttName, AttVal).
\end{verbatim}
and with the conjunctive query below that is equivalent to the equi-join above.
\begin{verbatim}
? res('Gene', Pk, 'GeneName', 'repA1'),
     res('Gene', Pk, 'Function', Val).
\end{verbatim}
If $k$ number of attributes of a concept were needed to be computed, we would be required to write a $k$-way conjunctive query in this approach.

Obviously, this query will fail. But, had the query been,
\begin{verbatim}
? res('Gene', Pk, 'GeneName', 'repA1'),
     res('Gene', Pk, 'UniProtProteinID', Val).
\end{verbatim}
it would have succeeded producing a binding {\tt O85067} for {\em Val}. It failed because the concept {\em gene} does not have a {\em function} in the base table {\em Entrez}, and consequently in the table {\em CanDB}. Therefore, the response remains empty unless CR tries an alternative evaluation.

\subsection{Indirect Response Generation}

The query above failed because concept {\em gene} does not have a property called {\em function}. Biologically, we know that genes encode proteins, and thus gene functions are manifest as protein functions, and thus are synonymous. Such knowledge are captured in the table {\em der} in figure \ref{ra}. A transitive closure of this relationship captures what we can compute as substitutes for a given concept, as captured in the axioms in rules 3 and 4 below.
\begin{verbatim}
2: res(Con, Pk, AttName, AttVal) :-
      rel(Con, Der, Pk, PkD),
      res(Der, PkD, AttName, AttVal).

3: rel(Con, Der, PkC, PkD) :- der(Con, Der),
      sO(Con, TabC, KeyC, AttsC),
      member(ColC, AttsC),
      sO(Der, TabD, KeyD, AttsD),
      member(ColD, AttsD),
      forK(TabC, TabD, ColC, ColD),
      canDB(Con, PkC, ColC, ValC),
      canDB(Der, PkD, ColD, ValC).
4: rel(Con, Der, PkC, PkD) :-
      rel(Con, DerC, PkC, PkDC),
      sO(DerC, TabC, KeyC, AttsC),
      member(ColC, AttsC),
      sO(Der, TabD, KeyD, AttsD),
      member(ColD, AttsD),
      forK(TabC, TabD, ColC, ColD),
      canDB(DerC, PkDC, ColC, ValC),
      canDB(Der, PkD, ColD, ValC).
5: member(Mem, [Mem|_]).
6: member(Mem, [_|Tail]) :- member(Mem, Tail).
\end{verbatim}
To reconstruct the extended object, for the property pair $\langle${\em function, AttributeValue}$\rangle$ from the corresponding concept {\tt Protein}, we need to establish the fact that {\tt Protein} indeed is a substitute (in the {\em der} table in figure \ref{ra}), {\em UniProtProteinID} {\tt O85067} corresponds to {\em GeneID} {\tt 1246500} (in table {\em Entrez} in figure \ref{tb}), and that the {\em Protein} {\tt O85067} has the pair $\langle${\em Function, AttributeValue}$\rangle$ in table {\em CanDB}, by virtue of table {\em UniProt} in figure \ref{ta}. We do so in rule 2 above by asserting a transitive relationship between the concepts {\tt Gene} and {\tt Protein}, and making sure they are connected by a foreign key relationship.

The rules 3 and 4 are a bit involved, but are conceptually simpler. These rules basically say, two objects are connected by a derived relationship is they have a direct foreign key relationship (rule 3), or a transitive foreign key relationship between the concept $C$ and $S$ such that $C$ has a direct foreign key relationship with some concept $D$ in {\em der} table and a foreign key relationship between the derived concept $D$ and some other concept $S$. The rules 5 and 6 are necessary axioms to test list memberships used in rules 3 and 4 to inspect the base tables have the required attributes. Adding rules 2 through 6 will now help us evaluate the subquery
\begin{verbatim}
   res('Gene', Pk, 'Function', Val)
\end{verbatim}
to be true with a binding {\tt Plasmid maintenance} for the variable {\em Val}, essentially computing the {\em Function} for the {\tt gene} {\tt repA1}, via {\em ProteinID} {\tt O85067}, and {\em GeneID} {\tt 1246500}.

\subsection{Interpretive Queries}

The indirect responses in the previous section are actually a class of queries that use conceptual substitutions to derive responses, i.e., property of proteins for genes know that they are substitutable as captured in the table {\em der} as a derivative. In fact any such biological knowledge can be encoded in BioSmart by creating a new rule for the predicate {\em rel/4} to link objects or concepts through their identifying keys. The independence of the {\em rel/4} predicate from how the antecedent is represented, makes it possible to make the mapping algorithm abstract and generic. For example, Scandinavian males, especially 33\%-45\% Swedish males, carry the I1 haplotype. We could then link genes of an offspring to his geographic origin, or to someone in that region, to discover possible properties. In such cases too, the predicate {\em rel/4} can be leveraged to link seemingly unrelated pieces of information to derive knowledge.

\begin{figure}[h]
    \centering{\includegraphics[width=.475\textwidth, height = 2.25in]{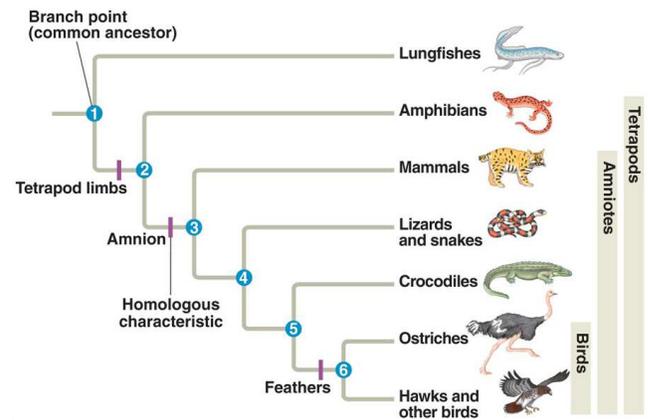}}
    \caption{Example phylogenetic tree.}
    \label{phylo}
\end{figure}

But such relationships are required to be in one of the database tables as ground values. In other words, no new information is actually generated, they are only linked in a meaningful and informative way. In BioSmart, we do allow a third kind of queries that actually allows generation of new knowledge not available in the database as ground facts using computational tools or functions. Applications of such tools have the potential to reveal new relationship among the database objects previously unknown, or discover new properties of objects in it. As shown in figure \ref{phylo}, species, as well their genes and morphologies, are related via genetic and morphological homology such as orthologs, paralogs, and horizontal gene transfer.

Aside from storing base facts in tables such as in PhylomeDB \cite{Huerta-CepasCPMG14} or TreeBASE \cite{TreeBASEII}, we can actually apply computational tools to compute these relationships and infer new properties. For example, orthologs can be computed using BLAST like tools such as Ortholog-Finder \cite{HoriikeMMNT2016}, ORCAN \cite{ZielezinskiDSK2017} or GENCODE \cite{Pei2012}, and phylogeny construction tools such as MEGA \cite{MEGA}, PAUP \cite{PAUP} or PHYLIP \cite{AllisonZ1999}. New relationships can also be established using databases such as GeneCards \cite{Safran2010}, and MapBase using ID mapping. In BioSmart, we allow such tool application based on background knowledge to link objects and then use the direct and indirect construction of properties using discovered relationships. For example, for the gene {\em repA1}, there are several orthologs that can be computed or retrieved from Entrez database or by using tools such as Ortholog-Finder or ORCAN: Gene ID {\tt 327491} in zebra fish, Gene ID {\tt 68275} in mouse, and Gene ID {\tt 417563} in chicken, each one of which could be used to decipher properties for the gene {\tt repA1}. It is interesting to note that recent studies show that although {\tt repA1} does not have paralogs in animals, it has paralogs in plants: {\tt RPA1A} to {\tt RPA1E} in Arabidopsis thaliana, and {\tt RPA1B} in Glycin max,  {\tt RPA1C} in Sorghum bicolor, and so on \cite{AkliluC2016}. This result is only available as of today in scientific literature, and not in any database, indicating that a text mining tool such as GeneView \cite{Thomas01072012} may be appropriate to discover this knowledge.

Calling such external functions from logic based languages such as Datalog, Prolog or XSB, is application specific. In general, they are called {\em foreign} codes or functions, and have specific protocols for implementation. In BioSmart, we use XSB \cite{Rao97} as a reasoning platform for its set based processing strategy, and Interprolog \cite{Miguel2004, interprolog} Java API for procedure calls. Interface {\em javaMessage()} of Interprolog binds XSB predicates with Java procedures. The rule implementing interpretive queries thus take the form below.
\begin{verbatim}
7: res(Con, PkC, AttNameR, AttValR) :-
      res(Con, PkC, _, _), simCon(Con, RelCon, Reln),
      cTool(RelnO, Ops), member(RelnO, Reln),
      res(RelCon, PkR, AttNameR, AttValR),
      member(Op, Ops), applyOp(Op, PkC, PkR).
\end{verbatim}
The rule 7 above basically asserts that we can derive the pair $\langle${\em AttNameR, AttValR}$\rangle$ for a object {\em Con} with a identifier {\em PkC}, if it is related to another object via the relationship in {\em simCon} table and an external call to the tool {\em Op} can verify the stated relationship with object {\em RelCon} with identifier {\em PkR} having the pair $\langle${\em AttNameR, AttValR}$\rangle$ as its property.

\subsection{Computational Tool Integration with XSB}
\label{XSB}

Predicate \textbf{applyOp()} in rule 7 above is a foreign function for XSB which is implemented procedurally in Java. It invokes different types of functions depending on the argument {\em Op}, i.e., it may access databases such as PhylomeDB, TreeBase, GeneCards or MapBase, it may initiate a computation by running desktop tools such as MEGA, PHYLIP, or PAUP, or look up the information in an web accessible tool such as WebPHYLIP or ORCAN. We use the \emph{Interprolog} \cite{interprolog} API to switch between the XSB reasoner and the procedural Java environment. Interface {\em javaMessage()} of Interprolog binds XSB predicates with Java procedures. The general form of \emph{javaMessage()} interface is shown bellow.
\begin{verbatim}
javaMessage(Target, Result, Exception, MessageName,
   ArgList, NewArgList)
\end{verbatim}
Interface \emph{javaMessage()} synchronously calls a method of a Java object \emph{Target}, then waits for its \emph{Result}, catching any \emph{Exception} that may occur. \emph{ArgList} specifies the arguments necessary for the function call, which must be of the proper Java-compatible types. \emph{NewArgList} contains the same objects in {\em ArgList} reflecting possible state changes after the function has been processed. Here, \emph{MessageName} is the invoked method of object \emph{Target}.

A Java class called \emph{MessageCaller} has been implemented to model the functions of \emph{javaMessages} invoked by XSB. A method named \emph{performOperation()} of \emph{MessageCaller} calls online tools such as BLAST, ORCAN, WebPHYLIP, or GENCODE based on the operations specified (sent as the argument list \emph{ArgList} by \emph{javaMessage}). For example, to compute homology of genes by the online tool BLAST, method \emph{performOperation()} uses the \emph{NCBI BLAST Java Interface} \cite{BLAST:Interface}. Given the ID of a gene, BLAST returns the IDs of the orthologs as an XML file. We then parse the XML file to retrieve the \emph{GeneIds} of the orthologous genes and send them back to XSB as \emph{Result}. An alternative technique for sending results is to store them in a database table and use an XSB predicate to retrieve those results within the XSB reasoner.

To be precise, the  \textbf{applyOp}({\em Op, PkC, PkR}) call in XSB is handled by the \emph{javaMessage()} operation as follows. XSB initiates the \emph{javaMessage()} call with the instantiations \emph{Target} to \emph{MessageCaller}, {\em Result} to {\em PkR}, \emph{MassageName} to \emph{performOperation()}, and \emph{ArgList} to $\langle$\emph{Op, PkC}$\rangle$. {\em Exception} and {\em NewArgList} are returned by Java as appropriate. Note that, the argument {\em Op} depends on the adornment of the variable by XSB before the call and depends on the tool list {\em Ops} in the {\em cTool} predicate. Eventually, method \emph{performOperation()} of class \emph{MessageCaller} applies the appropriate tool to perform the intended operation expected in the \textbf{applyOp}({\em Op, PkC, PkR}) predicate.

\section{Accessing Online Tools and \\ Databases}
\label{access}

In BioSmart, we leverage another level of abstraction for the interpretive class queries that opens up the possibility  endless ways computing them, and infer knowledge in unprecedented ways. That also means the Java \emph{MessageCaller} will need to be implemented on a case by case, which we believe is a daunting task. To avoid such unique implementation for each call type, we have decided to use the power of the abstraction encoded into the BioFlow language introduced earlier. This language, and its variants, have been leveraged in our implementation of LifeDB, MapBase, VizBuilder and VizFlow. In BioFlow, tools, databases and online web interfaces are viewed as function calls and are abstracted uniformly. Therefore, depending on the invocation and the specific tool, it is capable of customizing the evaluation.

In most tool applications, database processing and web applications, some form of selection conditions are applied (input arguments to a function), and results are extracted (output of the operation) before and after the operations are performed. There is also some form of schema mismatch between the terminologies used in the XSB program and the target system. Without the abstraction, users and the Java application writers will need to be fully aware of these terminologies and resolve the disparities manually. The BioFlow syntax already includes the machineries needed for schema matching and wrapping, in addition to accessing remote sites. The statement to access deep web resources in BioFlow is the {\sf extract} statement with the following syntax,
\begin{quote}
	{\sf extract} $A_1,\ldots, A_k$\\
	{\sf using matcher} $\mu$ {\sf wrapper} $\omega$ {\sf filler} $\phi$\\
	{\sf from} $\varphi$
	{\sf submit} $r$
	{\sf where} $\theta$
\end{quote}
where $\theta$ is the form condition, $A_1, \ldots, A_k$ is the projection list, $\mu$ is the schema matcher (e.g., PruSM \cite{NguyenNF10}), $\omega$ is the wrapper (e.g., FastWrap \cite{FastWrap-ijseke-2010s}), $\phi$ is the form filler (e.g., iForm \cite{TodaCSM10}), and $\varphi$ is the web form address or the form function. Note that this statement returns a table by submitting columns from each row in $r$ to the deep web database at $\varphi$. When a matcher, wrapper or filler isn't necessary, the corresponding clauses can be omitted. This statement can be constructed using the stepwise mapping of a resource such as $d_1$. Note that, in the case of a web service at $\varphi$, a form filler and a wrapper are not needed as the web service itself handles these functions, and no form conditions ($\theta$) are necessary either. But a transformation function $\tau$ may be required to convert XML to flat list of attributes, which could be made available in BioSmart library and help BioFlow extract the target fields automatically with a statement similar to the one below.
\begin{quote}
	{\sf extract} $A_1,\ldots, A_k$\\
	{\sf using matcher} $\mu$ {\sf transformer} $\tau$\\
	{\sf from} $\varphi$
	{\sf submit} $r$
\end{quote}
Analogously, when the text data are delimited in some way and are flat, we can expect a statement similar to the one below when inputs are not needed and hence the {\sf submit} clause can be omitted, but a wrapper $\omega$ is required to be able to read the data.
\begin{quote}
	{\sf extract} $A_1,\ldots, A_k$\\
	{\sf using matcher} $\mu$ {\sf wrapper} $\omega$\\
	{\sf from} $\varphi$
\end{quote}
These features of BioFlow allow users to write applications without having to worry about the details of extraction methods, location, technology specific nuances, format, and schema heterogeneity. The adoption of BioFlow also allows us to implement complex workflows using the applications similar to VisFlow. Users now can conceptualize a complete workflow at the highest abstraction level keeping a global scheme in mind knowing that the underlying data management and integration apparatus will be able to map her application onto potentially heterogeneous resources correctly and efficiently without any loss in query semantics.

For the call \textbf{applyOp}({\em BLAST, repA1, PkR}), we will construct the following BioFlow statement to execute.
\begin{quote}
	{\sf extract} {\em GeneID}\\
	{\sf using matcher} {\em PruSM} {\sf wrapper} {\em FastWrap} {\sf filler} {\em iForm}\\
	{\sf from} {\em 'http://blast.ncbi.nlm.nih.gov/Blast.cgi'}\\
	{\sf submit} {\em ArgRel}
\end{quote}
In the above expression, we will supply {\tt repA1} as the lone tuple in the table {\em ArgRel}. In fact, BioFlow can process set of inputs and thus is capable of a set based processing which can be utilized to speed up XSB predicate evaluation using external functions.

\section{Summary and Future Research}
\label{conclusion}

While there is a great deal of opportunities and interests in non-traditional data management and querying using key word based, NoSQL or natural language over unstructured databases, querying of structured databases using these approaches are also equally interesting. In this paper, and in our earlier research, we have demonstrated that interesting knowledge rich queries can be answered using such approaches, especially in investigative applications. Our contention is that users need not go to unimaginable lengths to dig out information just because they did not know how to. The BioSmart system we propose demonstrates the opportunities that exist and what is possible.

The concept reasoner, and domain specific knowledgebase we have leveraged is one of the major tool boxes that make BioSmart actually smart. But designing these components are manual, often application specific and tedious. But it also makes it possible to apply BioSmart to other scientific domains just by changing these components. Opportunities exists to use ontologies such as GO and SnoMed CT \cite{SNOMED-CT,WassermanW03s} to help users frame effective natural language queries by allowing terminology independence. This will require mapping mapping query terms to conceptual terms in the ontologies and use the standardized terms in the mapped queries. We are investigating an approach designing the CR and DSK components, at least partially, automatically from these components.

In Schema-Free SQL \cite{LiPJ14s}, meta information such as relation trees were used to aid query processors to tolerate mismatch or errors in schema information, and in BioVis \cite{Jamil14s} and ConLog \cite{ConLog-bmcgenomics-2012} conceptual structures were leveraged to map queries properly to underlying scheme. In BioSmart the need for such a structure to appropriately select parts of schema to frame the queries is much greater. Currently, the engineering of this structure is application specific and manual. Developing a similar conceptual structure generation scheme at least semi-autonomously will significantly improve usability of BioSmart. These are some of the issues we plan to continue as our future research.

\bibliographystyle{abbrv}

\end{document}